\documentclass[conference]{IEEEtran}
\IEEEoverridecommandlockouts
\usepackage{cite}
\usepackage[inline]{enumitem}
\usepackage{amsmath,amssymb,amsfonts}
\usepackage{algorithmic}
\usepackage{graphicx}
\usepackage{textcomp}
\usepackage{xcolor}
\usepackage{subfig}
\def\BibTeX{{\rm B\kern-.05em{\sc i\kern-.025em b}\kern-.08em
    T\kern-.1667em\lower.7ex\hbox{E}\kern-.125emX}}
\begin{document}

\title{TorKameleon: Improving Tor’s Censorship Resistance with K-anonymization and Media-based Covert Channels}

\author{\IEEEauthorblockN{Afonso Vilalonga}
\IEEEauthorblockA{\textit{NOVA LINCS \& DI,} \\
\textit{FCT, Universidade NOVA de Lisboa}\\
j.vilalonga@campus.fct.unl.pt}
\and
\IEEEauthorblockN{João S. Resende}
\IEEEauthorblockA{\textit{NOVA LINCS \& DI,} \\
\textit{Universidade do Porto}\\ 
jresende@fc.up.pt}
\and
\IEEEauthorblockN{Henrique Domingos}
\IEEEauthorblockA{\textit{NOVA LINCS \& DI,} \\
\textit{FCT, Universidade NOVA de Lisboa}\\
hj@fct.unl.pt}
}

\maketitle

\begin{abstract}
Anonymity networks like Tor significantly enhance online privacy but are vulnerable to correlation attacks by state-level adversaries. While covert channels encapsulated in media protocols, particularly WebRTC-based encapsulation, have demonstrated effectiveness against passive traffic correlation attacks, their resilience against active correlation attacks remains unexplored, and their compatibility with Tor has been limited. This paper introduces TorKameleon, a censorship evasion solution designed to protect Tor users from both passive and active correlation attacks. TorKameleon employs K-anonymization techniques to fragment and reroute traffic through multiple TorKameleon proxies, while also utilizing covert WebRTC-based channels or TLS tunnels to encapsulate user traffic.
\end{abstract}


\section{Introduction}
Tor is a widely used anonymization network that operates based on the Onion Routing protocol~\cite{tor}. Its primary objective is to provide low-latency communication while preserving user anonymity. This is achieved by establishing network paths called Tor circuits, which typically consist of three nodes known as Tor relays. Routing traffic through the Tor circuits decouples the user's incoming and outgoing traffic, theoretically rendering them unlinkable and reducing the likelihood of an attacker correlating the two, thereby preserving user anonymity with regard to what they are accessing online. However, in practice, Tor exposes vulnerabilities, particularly in the context of deanonymization attacks~\cite{tor_attacks,DeepCoFFEA,deepcorr,finn,FlowTracker,tormarker,Effective_Detection_ML}. Studies have demonstrated the effectiveness of statistical analysis, machine learning, and deep learning models in identifying similarities between incoming and outgoing Tor network flows, making many Tor circuits susceptible to passive correlation attacks~\cite{AS-level_tor}. This concern is magnified when considering authoritarian regimes and states with the resources to control extensive autonomous systems (ASs) and deploy large-scale censorship mechanisms and apparatus that use such deanonymization attacks.

The Tor Project has developed techniques to mitigate such vulnerabilities, primarily by introducing pluggable transports into the network. Pluggable transports are software components that mask the Tor traffic exchanged between the user and the Tor entry relay of the circuit (i.e., the Tor Bridge) by randomizing it, encapsulating it, or employing other traffic obfuscation techniques. By modifying or concealing the characteristics of incoming traffic, the expectation is that any attempt to correlate it with outgoing traffic will be rendered ineffective. While numerous pluggable transports have been developed and continue to undergo refinement, it is important to emphasize that even the most well-known and widely used ones have shown vulnerability to correlation attacks and other deanonymization methods~\cite{faccumul,snowflake_attack,the_parrot_is_dead,attacks_aginst_meek_and_obsf}.

These vulnerabilities have prompted researchers to explore alternative systems resistant to correlation attacks and develop new solutions for evading internet censorship. One promising research direction involves traffic encapsulation, where traffic is concealed within a covert carrier, such as another protocol. Several solutions have emerged using media protocols~\cite{freewave,facet_art,CoverCast,deltashaper}, with two advanced systems employing WebRTC as their protocol carrier~\cite{protozoaPaper,stegozoa}. The widespread adoption and dissemination of WebRTC make it inconspicuous and less likely to be blocked by censors due to potential collateral repercussions. These two WebRTC-based censorship evasion systems, Protozoa and Stegozoa, have demonstrated resilience against passive correlation attacks. However, they face challenges related to integration with the Tor network, deployment complexity, and the need for testing against the growing and potent trend of active correlation attacks employed by censors, aiming to inject watermarks like temporal delays into traffic at specific network segments with the aim of propagating throughout the network and becoming detectable elsewhere.

In this paper, our goals are simple: \begin{enumerate*}[label = \Roman*)] \item To test the effectiveness of WebRTC in resisting state-of-the-art active correlation attacks; \item To develop an evasion system capable of resisting both active and passive state-of-the-art correlation attacks employed by current censors while maintaining reasonable performance for low-throughput Internet tasks; \item To ensure full compatibility of the system with the Tor network by developing it as a pluggable transport.\end{enumerate*} To achieve these goals, we have developed TorKameleon, a Tor pluggable transport that can encapsulate Tor traffic within WebRTC video conference streams and TLS tunnels, enabling a greater diversity of traffic flowing through the network. 
Furthermore, it enables the creation of a pre-Tor network consisting of $K$ TorKameleon proxies. This network facilitates the fragmentation and routing of user traffic among the proxies, employing a multipath strategy that blends user traffic with that of $K$ other users, a technique referred to as K-anonymization. TorKameleon can withstand passive and active correlation attacks simulating the deanonymization efforts of a state-level adversary, all while maintaining reasonable throughput for low-throughput Internet tasks. The contributions of this work can be summarized as follows:  \begin{enumerate*}[label = \Roman*)] \item A complete specification of the TorKameleon system; \item An implementation of the designed solution available as an open-source prototype~\cite{github}; \item A comprehensive experimental evaluation of the system, assessing its performance and unobservability against active and passive correlation attacks. \end{enumerate*}

The rest of this article is structured as follows: Section~\ref{related_work} presents related work. Section~\ref{system} details the TorKameleon system model. Section~\ref{experimental evaluation} focuses on preliminary performance and correlation attack resistance evaluation. Finally, Section~\ref{conclusion} summarizes findings and discusses future research directions.

\section{Background and Related Work}
\label{related_work}
Over the years, research has emphasized anonymization systems to evade censorship and counter deanonymization attacks, resulting in the development of various techniques. This section explores correlation attacks and their countermeasures, such as media-based protocols and K-anonymization systems.

\subsection{Correlation Attacks}
Correlation attacks are techniques used to extract information and create user profiles of specific targets or deanonymize communicating endpoints within a network~\cite{DeepCoFFEA,deepcorr,finn,FlowTracker,tormarker,Effective_Detection_ML}. These attacks can be executed by state-level adversaries who control multiple AS regions and collaborate with organizations like internet service providers (ISPs). In the context of Tor, an attacker controlling both the entry and the exit Tor relays in a circuit will attempt to correlate inbound and outbound traffic to identify which pairs of flows belong to the same overall flow. To achieve this, the attacker analyzes metadata such as inter-packet arrival times, packet lengths, and volumes~\cite{Effective_Detection_ML}. Using this information, attackers can confirm with a high degree of probability that a particular user is accessing a specific web service. Passive correlation attacks~\cite{DeepCoFFEA,deepcorr,Effective_Detection_ML,FlowTracker} involve passively observing and collecting traffic to later correlate it and deanonymize the target, while active correlation attacks~\cite{finn,tormarker} inject a watermark or fingerprint into packets with the hope that it propagates through the network and becomes observable elsewhere, uniquely identifying the traffic flow's origin or that the flow was watermarked. This watermark or fingerprint consists of a recognizable pattern inserted into the traffic as it passes through a specific point in the network, such as a temporal delay.

\subsection{Multipath and K-Anonymization}
The principle of K-anonymization originated as a method to anonymize database records~\cite{k-anonymization} and has since been extended to various domains with the aim of reducing the probability of correct identification by attackers to at most $1/K$. TorK~\cite{tork} and Tir~\cite{tir} are K-anonymization systems designed for Tor traffic. Tir also enables the utilization of a multipath routing and traffic fragmentation strategy across a network of $K$ proxies. They achieve high throughputs (12 Mbps for TorK and 1.6 Mbps for Tir) while remaining undetected against passive correlation attacks. However, these systems have only undergone testing against passive correlation attacks. Additionally, using TorK and Tir can be complex, as they necessitate maintaining a constant group of users to preserve the $K$ set, with potential failures that could result in information leaks through traffic analysis.

\subsection{Media Protocol Tunneling}
Media tunneling solutions utilize media protocols such as video and audio streams to embed covert data. These solutions provide improved alternatives for bypassing censorship. FreeWave~\cite{freewave} is a circumvention tool that leverages audio signals from VoIP connections to tunnel hidden Internet traffic, achieving a throughput of 19.2 kbps. Facet~\cite{facet_art} enables users to watch videos by tunneling them through video conferences via Skype, offering a maximum throughput of 471 kbps. CovertCast~\cite{CoverCast} transforms website content into videos and streams them via popular platforms such as YouTube, offering a method to bypass censorship with a throughput of 168 kbps. DeltaShaper~\cite{deltashaper} facilitates the covert transmission of TCP/IP traffic by tunneling the traffic through Skype, achieving a throughput of 7 kbps. However, it is worth noting that CovertCast, Facet, and DeltaShaper can be detected with over 90\% accuracy, and FreeWave can be easily identified through audio signal analysis or intentional disruptions in communication~\cite{cover_your_pitfalls,Effective_Detection_ML}. WebRTC tunneling, the recent state-of-the-art in covert traffic media encapsulation, utilizes the WebRTC stack to establish covert channels. Protozoa~\cite{protozoaPaper}, a censorship evasion tool, employs WebRTC-based media streaming web apps for covert IP packet encapsulation, offering unobservability, unblockability, and high throughput (approximately 1.4 Mbps). Stegozoa~\cite{stegozoa}, an extension of Protozoa, employs steganography to hide data within video streams, ensuring unobservability even against malicious gateways, achieving a throughput of roughly 8.2 kbps. Snowflake~\cite{snowflake}, a Tor pluggable transport, communicates with temporary web browser-based proxies via WebRTC data channels, which are designed for arbitrary data transmission (unlike Protozoa and Stegozoa, which use the video streaming channel), achieving throughput ranging from 1.5 Mbps to 1.7 Mbps~\cite{snowflake_throughput}. However, it is susceptible to detection through traffic analysis~\cite{snowflake_attack, faccumul}. Protozoa and Stegozoa resist passive correlation attacks but have not been tested against active correlation attacks. Furthermore, deploying them can pose challenges, necessitating users to compile the Chromium browser, and they lack compatibility with Tor

\subsection{TorKameleon Comparison}
TorKameleon is an innovative censorship evasion system that leverages the strengths of K-anonymization and Web-RTC media tunneling. Notably, it enhances Tor's resistance against deanonymization attacks by functioning as a fully Tor-compatible pluggable transport. To the best of our knowledge, in contrast to the previously described systems, TorKameleon has undergone rigorous testing against active correlation attacks, enabling the assessment of WebRTC encapsulation's resilience against these attacks. While providing performance comparable to the previously mentioned systems, we offer a highly configurable solution that is fully compatible with Tor and resilient against both active and passive correlation attacks, making it easily deployable in real-world scenarios.

\section{TorKameleon}
\label{system}
In this section, we provide an overview of the inner workings of the TorKameleon system and discuss the assumed threat model.

\subsection{System Model}
\begin{figure}[ht] 
    \centering
    \includegraphics[width=0.9\linewidth]{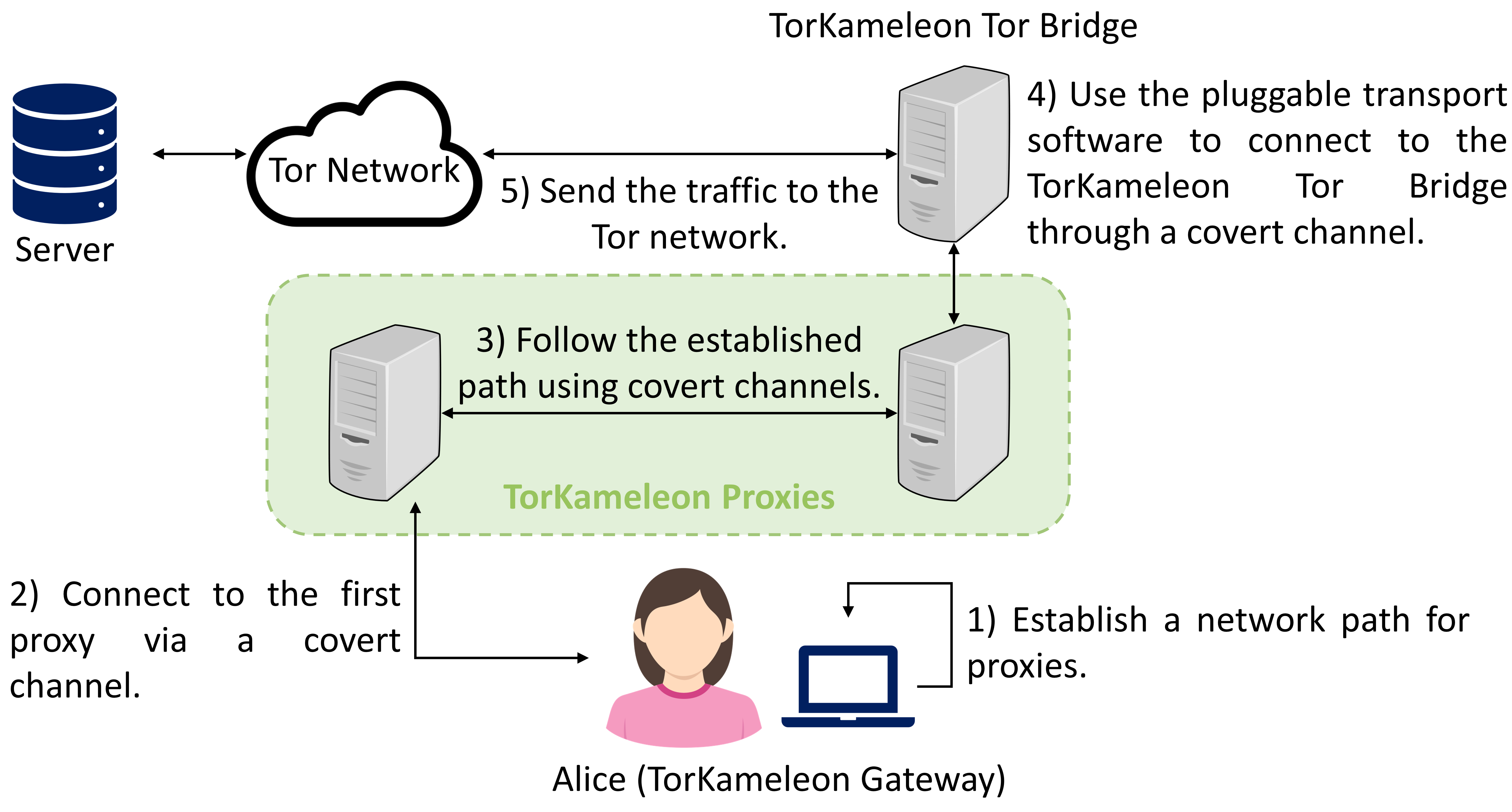}
    \caption{System Model and Workflow of the TorKameleon Ecosystem. When using the pluggable transport without proxies, the user establishes a direct connection to the TorKameleon Tor Bridge through the TorKameleon pluggable transport client-side, which operates on the user's local device.}
    \label{system model} 
\end{figure}

The TorKameleon system operates in three modes: \begin{enumerate*}[label = \Roman*)] \item \textbf{Pluggable Transport Mode}: Users connect to the Tor Bridge via the TorKameleon pluggable transport, exchanging Tor traffic encapsulated in WebRTC video conferences or TLS tunnels; \item \textbf{Standalone Mode}: Users, either individually or in groups, deploy TorKameleon proxies to fragment and route user traffic across multiple paths composed of multiple proxies using encapsulation channels; \item \textbf{Combined Mode}: User traffic is routed through the proxy network before being sent to the Tor network via the pluggable transport. \end{enumerate*} We provide an overview of the system in Figure~\ref{system model}.

Consider Alice in a censored region. Alice begins by configuring a network path of proxies or letting the TorKameleon gateway software (our local client software) determine the path. Next, Alice connects to the first TorKameleon proxy using the TorKameleon gateway, either through a TLS tunnel or a covert WebRTC channel in a video conference. Alice's traffic can now be routed through covert connections to other proxies in the TorKameleon network. These covert connections also use TLS tunneling or WebRTC-based channels. Alternatively, Alice's traffic can go directly to the Tor network via our TorKameleon Tor Bridge. Each TorKameleon proxy runs the Tor daemon and TorKameleon pluggable transport client software locally. When Alice's traffic is intended for the Tor network, the proxy forwards it to the local Tor daemon. From there, it is sent to the TorKameleon client-side pluggable transport, which then transmits it to the connected TorKameleon Tor Bridge via a covert channel. Upon arrival at the TorKameleon Tor Bridge, the traffic is decapsulated and subsequently forwarded into the Tor network.

\subsection{Threat Model}
We consider a state-level adversary with the capability to collaborate with entities such as ISPs and other governments. The primary objective of the censor is to identify and block the use of TorKameleon while minimizing any impact on legitimate WebRTC and TLS connections. The censor can monitor, capture, and analyze all network traffic originating from the user, TorKameleon proxies, TorKameleon bridges, and the Tor network, as long as the network segments accessed fall within its jurisdiction or that of the involved adversary parties. We also assume that the censor can employ active correlation attacks, utilizing watermarking. However, the software installed on users' devices, TorKameleon Tor bridges, and proxies is assumed to be uncompromised, and no intrusive actions are performed within these systems. Additionally, we assume that the censor does not engage in widespread blocking or disruption of WebRTC traffic or TLS communications. Such actions would cause significant collateral damage and negatively impact legitimate users and services that rely on these protocols. Furthermore, we assume that the adversary does not have the capability to operate controlled TorKameleon nodes and does not have access to the clear video stream (i.e., unencrypted video stream).

\subsection{System Architecture}
TorKameleon consists of two subsystems that operate similarly: the TorKameleon proxy and the TorKameleon pluggable transport. The TorKameleon pluggable transport is a bundle of software comprising the client-side pluggable transport (cs-pt) and the server-side pluggable transport (ss-pt). Both systems consist of five main components: the WebRTC encoder and decoder, the TLS tunnel, the SOCKS5 proxy, and the controller. These components work seamlessly together, with minor differences between the proxy and the pluggable transport, across three operational stages: the covert channel establishment stage, the encapsulation stage, and the networking stage (similar to both Protozoa and Stegozoa).

\paragraph{\textbf{Covert Channel Establishment Stage}}The establishment of the covert channel plays a pivotal role within the TorKameleon system. Depending on the system configuration, the controller establishes either a TLS tunnel or a WebRTC covert channel. For the TLS tunnel, a messaging protocol to establish an SSL Tunnel is required. However, for the WebRTC covert channel, the WebRTC-based video conferencing application must be initialized, including the signaling protocol. This allows the TorKameleon proxies, or the cs-pt and ss-pt, to establish a video conference and consequently send encapsulated data (further details in Section~\ref{WebRTC enc}). In the case of the proxy, the controller is also responsible for managing the proxies chosen for the multipath circuits.

\paragraph{\textbf{Encapsulation Stage}} The TLS and WebRTC encapsulation components are critical elements of the TorKameleon system, performing essential tasks related to encapsulating and decapsulating covert traffic. The TLS component is responsible for managing and routing traffic to the TLS tunnel through the Stunnel system. The WebRTC encoder is tasked with encoding user or Tor traffic into WebRTC video frames, while the WebRTC decoder handles the decoding of encapsulated traffic. Packets to be encoded are placed in a queue until a frame is available for encoding. The WebRTC encoding mechanism includes fragmentation and reassembly mechanisms required for the proper reordering of encapsulated packets.

\paragraph{\textbf{Networking Stage}}Regarding the networking stage of the TorKameleon pluggable transport, the Tor daemon transmits Tor traffic to the cs-pt (which runs on the user's device or the TorKameleon proxy) through a SOCKS5 proxy. On the other hand, the ss-pt (which operates on the Tor bridge) employs a reverse proxy to establish a connection with the Tor daemon, enabling the routing of traffic throughout the rest of the Tor network. The TorKameleon proxy uses default sockets to establish connections between user applications and the locally running proxy software (the TorKameleon gateway). Additionally, support for the SOCKS5 protocol between the application and the TorKameleon gateway is also provided.

\subsection{WebRTC-based Encapsulation}
\label{WebRTC enc}
To encapsulate traffic within a WebRTC video conference stream, we have created a browser-based video conferencing web app using the WebRTC technology stack. This app utilizes the widely supported WebRTC API found in most browsers, enabling video conferencing between two participants. Using the video stream as a channel for transmitting encapsulated data, the WebRTC encapsulation component conceals hidden traffic within the seemingly innocuous video traffic. This process involves two main steps: web application initialization and data encapsulation.

\paragraph{\textbf{Web Application Initialization}} We employ Node.js servers to locally serve the code necessary to run the WebRTC web application. Within our WebRTC application, two roles exist: the initiator, responsible for initiating the conference call, and the receiver, which awaits the initiator's call. For instance, when the cs-pt intends to establish a covert channel with the ss-pt, the cs-pt controller component instructs the ss-pt controller component to initiate the WebRTC web application in receiver mode through a messaging protocol. After confirmation, the cs-pt starts its WebRTC web application in initiator mode. The signaling process then occurs, leading to the establishment of the video conference. To automate browser functions and access the WebRTC web application, we have utilized the Selenium framework~\cite{selenium}.

\begin{figure}[ht] 
    \centering
    \includegraphics[width=0.90\linewidth]{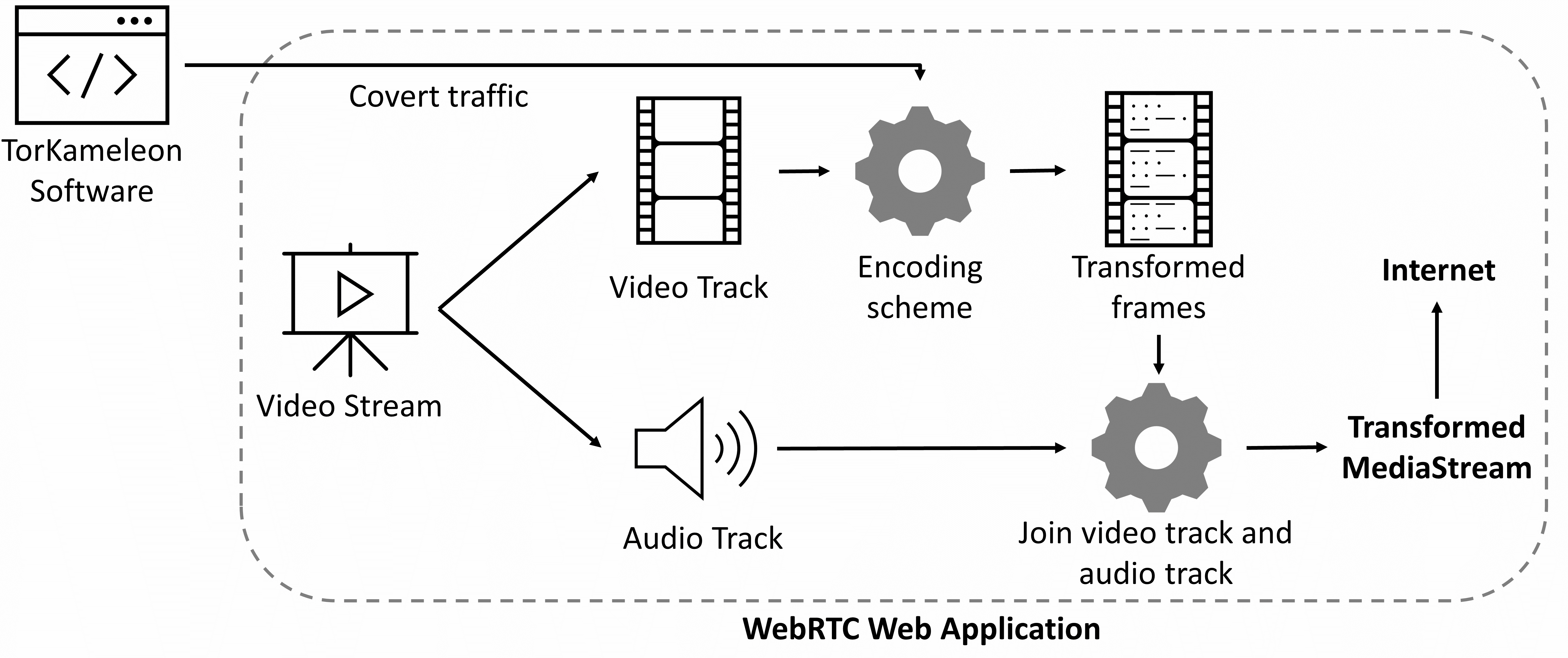}
    \caption{Workflow for encapsulating user traffic in WebRTC video frames.}
    \label{workflow_enc} 
\end{figure}

\paragraph{\textbf{Data Encapsulation}} In Figure~\ref{workflow_enc}, we outline the WebRTC encapsulation process. It begins with the split of the video stream into audio and video tracks, followed by the insertion of the covert data. This data is obtained via WebSockets between the TorKameleon software and the WebRTC web application. The modified video track, now containing encapsulated data, is then combined with the audio track to create a new media stream. This traffic appears as regular WebRTC traffic, effectively disguising its true nature from censors. The new stream is subsequently sent to the other web conference peer for later decapsulation. This is accomplished using the WebRTC Insertable Streams API, which enables the manipulation of video and audio frames.

When the web application receives an array of $X$ bytes of data to be encapsulated from TorKameleon (with $X$ being user-configurable), it stores this array and waits for a video frame to encapsulate the data. This array is referred to as a ``data block''. Two modes exist for embedding data blocks into frames: \texttt{ADD} and \texttt{REPLACE}. In \texttt{ADD} mode, the data block is appended to the frame without altering the existing content. It integrates the entire data block into a single packet, attached to a single video frame. In \texttt{REPLACE} mode, the data block replaces the frame content while preserving the frame header. This mode is more complex and may involve fragmenting the data block based on the available video frame size.

\section{Experimental Evaluation}
\label{experimental evaluation}
This section provides initial tests of TorKameleon, assessing its performance and resistance to passive and active correlation attacks. While ongoing, these early results showcase TorKameleon's capabilities.

\subsection{Setup}
Our setup included five machines running Ubuntu 20.04. Four of them were OVH virtual private servers (VPS) equipped with Intel 8-core CPUs running at 2.4 GHz, 32 GB of RAM, and a 2 Gbps bandwidth. The fifth system was a local one, featuring an Intel i5-9300H CPU with 4 cores running at 2.4 GHz, 16 GB of RAM, and a 1 Gbps bandwidth. Except for two, all were in different locations—France, UK, Canada, and Portugal. In tests using the Tor network, we used a fixed set of three circuit relays: the first in our control, the middle in Germany, and the exit relay in the Netherlands. For performance assessment, we evaluated both latency and throughput. To measure throughput, we downloaded a 250 KB file from an HTTP server running on the Canada VPS. Latency was measured using the httping tool, which recorded the time to receive the initial byte in response to an HTTP/HTTPS request. We tested with multiple users, peaking at 50 in parallel, corresponding to the daily Tor Bridge user numbers~\cite{tor_daily_users_bridge}. Each user operated within a Docker container. Latency results averaged ten measurements; throughput averaged five. Each test was repeated twice, with different data block sizes (536, 1050, 2078, and 4134 bytes), corresponding to 1, 2, 4, and 8 Tor cells (including header size). The results were then compared to default Tor performance metrics.

\subsection{Performance}
\begin{figure}[ht]
    \centering
    \subfloat[\label{1a}]{%
        \includegraphics[width=0.5\linewidth]{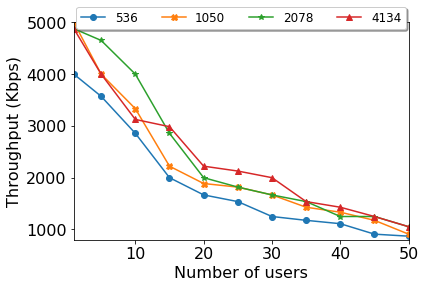}}
    \hfill
    \subfloat[\label{1b}]{%
        \includegraphics[width=0.5\linewidth]{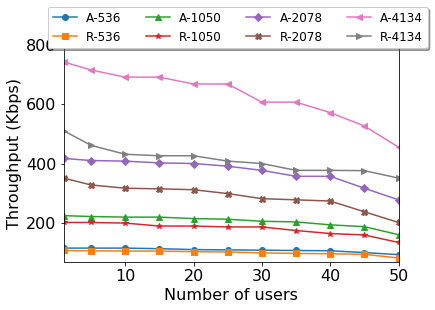}}
  \caption{Throughput graphs. (a): Throughput graph for TLS encapsulation. (b): Throughput graph for WebRTC encapsulation. A-\texttt{ADD} mode; R-\texttt{REPLACE} mode. \vspace{-10pt}}
  \label{throughput} 
\end{figure}

Figure~\ref{1a} illustrates the throughput of TorKameleon when utilized as a pluggable transport with TLS encapsulation. As depicted in the graph, minimal variation in throughput is observed across different data block sizes. The throughput values with only one user closely match those obtained in the Tor metrics website for throughput~\cite{tor_metrics} and in our baseline validation (5128 Kbps). Figure~\ref{1b} depicts the throughput when utilizing WebRTC-based encapsulation with 50 users incrementally added to the TorKameleon Tor Bridge. To manage load and prevent congestion, 5 clients utilize WebRTC-based encapsulation (including the one from which we took our measurements), while the remaining users employ TLS tunneling. The graph presents throughput values for both \texttt{ADD} and \texttt{REPLACE} modes across various data block sizes. The results show reasonable throughput for basic internet tasks, using WebRTC encapsulation, ranging from 108 Kbps (\texttt{REPLACE} mode with 536 bytes) to 741 Kbps (\texttt{ADD} mode with 4134 bytes). This is comparable to the related work mentioned in Section~\ref{related_work}. The reduction in throughput compared to the Tor default is attributed to limitations imposed by the available frames per second for encapsulation, the frame size for data replacement (particularly in \texttt{REPLACE} mode, which is also why \texttt{ADD} mode outperforms \texttt{REPLACE} mode), and the overhead of the encapsulation process.

\begin{figure}[ht] 
    \centering
    \subfloat[\label{2a}]{%
        \includegraphics[width=0.5\linewidth]{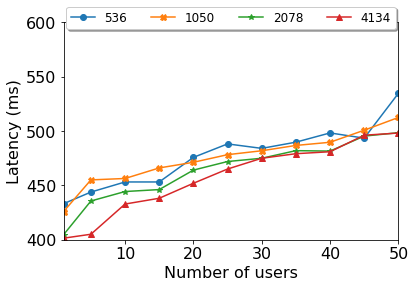}}
    \hfill
    \subfloat[\label{2b}]{%
        \includegraphics[width=0.5\linewidth]{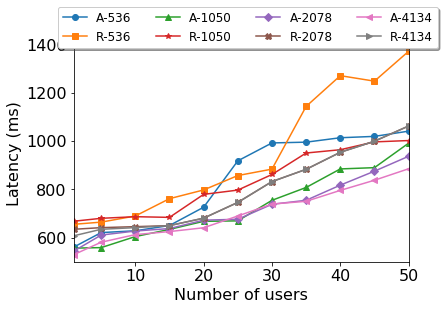}}
  \caption{Latency graphs. (a): Latency graph for TLS encapsulation. (b): Latency graph for WebRTC encapsulation. A-\texttt{ADD} mode; R-\texttt{REPLACE} mode.}
  \label{latency} 
\end{figure}

Figure~\ref{2a} shows latency values for the TLS encapsulation mode, while Figure~\ref{2b} displays latency values for the WebRTC-based encapsulation in both \texttt{ADD} and \texttt{REPLACE} modes. Latency remains consistent across different data block sizes, indicating similar performance. TLS latency values for a single user closely align with the default Tor latency metrics (400 ms)~\cite{tor_metrics}. However, WebRTC encapsulation, in both \texttt{ADD} and \texttt{REPLACE} modes, exhibits slightly higher latencies compared to the default Tor, ranging from 529 ms (\texttt{ADD} mode with 4134 bytes) to 655 ms (\texttt{REPLACE} mode with 536 bytes). These latency values do not notably affect the usability of TorKameleon for internet tasks like downloading files and regular web browsing.

\subsection{Resistance to Passive and Active Correlation Attacks}
\begin{figure}[ht]
    \centering
    \subfloat[\label{3a}]{%
        \includegraphics[width=0.5\linewidth]{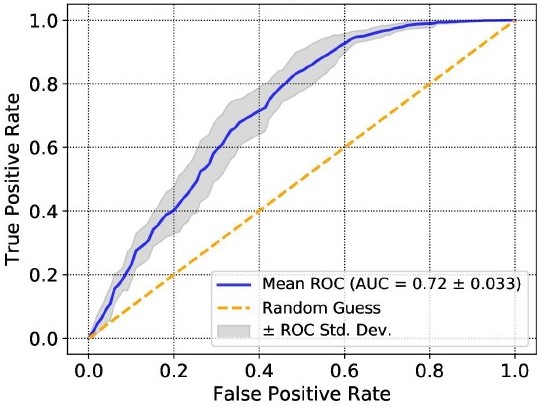}}
    \hfill
    \subfloat[\label{3b}]{%
        \includegraphics[width=0.5\linewidth]{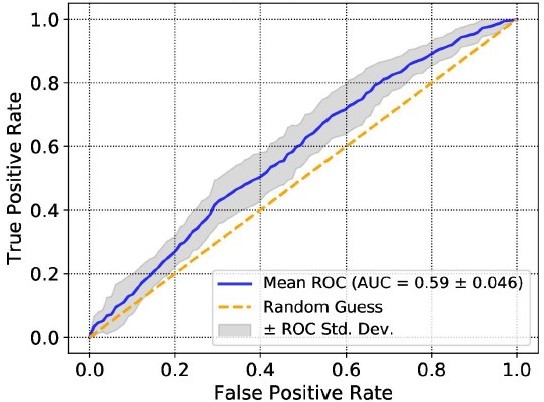}}
  \caption{Passive correlation attacks. (a) One proxy. (b) Four proxies. \vspace{-10pt}}
  \label{passive} 
\end{figure}

To evaluate passive correlation attack resistance, we employed the XGBoost classifier, consistent with prior state-of-the-art research~\cite{Effective_Detection_ML}. In Figure~\ref{passive}, we compare a single proxy to a network of four proxies, all connected via TLS tunnels, while observing only one proxy. We conclude that adding more proxies increases the available paths for traffic rerouting and reduces the volume of traffic passing through the monitored proxy (assuming the attacker can't access traffic from all deployed proxies due to their geographic distribution). As such, the TorKameleon system with a network of TorKameleon proxies alone can effectively resist passive correlation attacks, achieving an Area under the ROC Curve (AUC) of up to 0.59 with four proxies, akin to random guessing.

\begin{figure} 
    \centering
  \subfloat[\label{6a}]{%
       \includegraphics[width=0.5\linewidth]{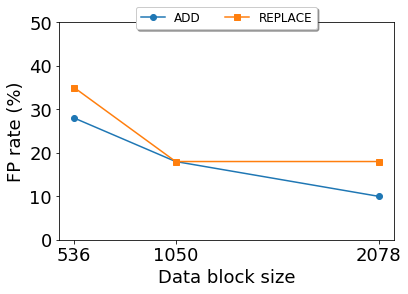}}
    \hfill
  \subfloat[\label{6b}]{%
        \includegraphics[width=0.5\linewidth]{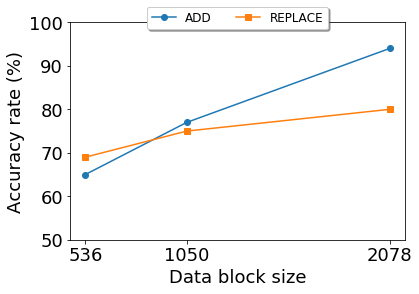}}
  \caption{Active correlation attack result graphs for WebRTC encapsulation in \texttt{ADD} and \texttt{REPLACE} mode with different data block sizes (bytes). (a) False Positive Rate (FPR) graph. (b) Accuracy rate graph. \vspace{-12pt}}
  \label{active} 
\end{figure}

We evaluated active correlation attacks using TorMarker~\cite{tormarker}, which 
injects watermarks into the traffic and uses deep learning to detect them. In this test, we exclusively utilize TorKameleon as a pluggable transport with WebRTC encapsulation. In Figure~\ref{active}, the results from the active correlation attacks reveal two key findings: smaller data block sizes enhance TorKameleon's unobservability, resulting in reduced accuracy and higher false positive rates (FPR); larger data block sizes make the system more vulnerable. Additionally, differences between \texttt{REPLACE} and \texttt{ADD} modes become more pronounced with larger data block sizes, with \texttt{REPLACE} mode proving more robust against watermarking attacks. TorKameleon maintains resistance against active correlation attacks when used with data blocks of sizes 536 and 1050 bytes, based on our defined thresholds of FPR not falling below 10\% and accuracy not exceeding 80\% 
(based on an analysis of experimental evaluations of other state-of-the-art tools~\cite{tormarker,deepcorr,DeepCoFFEA}).

\section{Conclusion}
\label{conclusion}
In this paper, we have introduced a novel censorship evasion tool aimed at addressing Tor's susceptibility to active and passive correlation attacks. TorKameleon utilizes multipath routing and traffic encapsulation via WebRTC media streams and TLS tunnels, and it is available on GitHub as an open-source project~\cite{github}. Furthermore, we demonstrated the results of WebRTC media encapsulation in the context of active correlation attacks, highlighting its potential to enhance anonymization systems like Tor. However, extending the current experimental evaluation is a critical next step.

\section*{Acknowledgment}
This work is supported by NOVA LINCS (UIDB/04516/2020) with the financial support of FCT.IP.

\end{document}